# Effects related to deposition temperature of ZnCoO films grown by Atomic Layer Deposition – uniformity of Co distribution, structural, optical, electrical and magnetic properties

Małgorzata I. Łukasiewicz [*,1], Bartłomiej Witkowski[1,2], Marek Godlewski[1,2], Elżbieta Guziewicz[1], Maciej Sawicki[1], Wojciech Paszkowicz[1], Rafał Jakieła[1], Tomasz A. Krajewski[1], Grzegorz Łuka

[1] *Institute of Physics, Polish Acad. of Sciences, Al. Lotników 32/46, 02-668 Warsaw, Poland*
[2] *Dept. of Mathematics and Natural Sciences College of Science Cardinal S. Wyszyński University, Dewajtis 5, 01-815 Warsaw, Poland*

In the present study we report on properties of ZnCoO films grown at relatively low temperature by the Atomic Layer Deposition, using two reactive organic zinc precursors (dimethylzinc and diethylzinc). The use of these precursors allowed us the significant reduction of a growth temperature to below 300°C. The influence of growth conditions on the Co distribution in ZnCoO films, their structure and magnetic properties was investigated using Secondary Ion Mass Spectroscopy, Scanning Electron Microscopy, Cathodoluminescence, Energy Dispersive X-ray Spectrometry (EDX), X-ray diffraction and Superconducting Quantum Interference Device magnetometry. We achieved high uniformity of the films grown at 160°C. Such films are paramagnetic. Films grown at 200° and at higher temperature are nonuniform. Formation of foreign phases in such films was detected using high resolution EDX method. These samples are not purely paramagnetic and show weak ferromagnetic response at low temperature.

[*] Corresponding author: e-mail mluk@ifpan.edu, Phone: +48 22 843 70 01, Fax: +48 22 843 09 26

## 1. Introduction

Diluted magnetic semiconductors (DMS) are potential materials for spintronic devices. One of the most studied DMS material is ZnO doped with 3d transition metal (TM) ions (Co, Mn, Ni, Fe, etc.) [1-11]. This is because theory of Dietl et al. [6] predicted that heavily p-type doped ZnO with 5% of Mn should be ferromagnetic (FM) at RT. Other calculations (now corrected by K. Sato during 5$^{th}$ Spintech 2009 conference) claimed possibility of RT FM in *n*-type ZnO with cobalt [4,5].

Concentrated research led to many conflicting reports on magnetic properties of ZnTMO samples. For example, several papers reported that ZnO exhibits ferromagnetism above room temperature (RT) when doped with few atomic percents of cobalt [1-5].

RT ferromagnetism of ZnCoO thin films was reported for materials grown with various growth methods, such as pulsed-laser deposition or sputtering (see e.g. [1,7,8]), but the reproducibility of such processes is questionable, it was is less than 10% in case of the results reported in the reference [1]. It is thus still controversial whether the observed ferromagnetism is intrinsic or comes from some secondary phases, as already discussed for another ZnO-based DMS material ZnMnO [9,10].

This paper presents and discusses the results of structural, optical, electrical and magnetic investigations of ZnCoO thin films obtained by the ALD method, the technique enabling relatively low growth temperature without scarifying the uniformity of the films.

## 2. Experimental

ZnCoO thin films were grown using the ALD technique, introduced by Suntola. We used organic precursors: dimethylzinc (DMZn) and diethylzinc (DEZn) as zinc precursors, cobalt (II) acetyloacetonate and cobalt (II) chloride hydrate as cobalt precursors. Deionized water was used as an oxygen precursor. These highly reactive precursors are sequentially introduced to the growth chamber, so they meet only at a surface of a growing film. The use of these precursors allowed us the significant reduction of a growth temperature to below 300ºC.

As substrate we used sapphire, GaAs, Si (for magnetic and structural measurements), sapphire (for optical investigations) and glass (for electrical experiments). ZnCoO samples were grown with different Zn-to-Co ratios of the ALD cycles, which turned out to be crucial to obtain uniform ZnCoO layers, as already indicated in the reference [12]. As-grown ZnCoO layers deposited on silicon were studied first and then were annealed at three different temperatures (at 400ºC, at 600ºC, and at 800ºC) in nitrogen atmosphere for two hours.

## 3. Results and discussion

Energy Dispersive X-ray Spectrometry (EDX) and Secondary Ions Mass Spectroscopy (SIMS) were used to measure the percentage of cobalt in ZnCoO films, and also an uniformity of cobalt distribution. The Co fractions obtained from both methods were similar and ranged between 1 to 6 atomic percents, depending on the growth temperature, Zn-to-Co ratio and length of precursor pulses in the ALD processes.

The structure of ZnCoO films was investigated with Scanning Electron Microscopy (SEM). Figures 1 (a-d) show the SEM micrographs of the $Zn_{0.98}Co_{0.02}O$ thin film for: (a) as-grown sample (grown at 160ºC), (b), (c) and (d) the same sample after annealing at 400°C (b), 600°C (c) and 800°C (d), respectively.

In most of the cases ZnCoO films showed rather uniform and smooth surface with granular microstructure coming from a columnar growth mode (observed in SEM experiments at cross-sections of samples). Diameter of the columns depended on the growth and annealing temperature. We observed that the growth and annealing temperature affects the growth mode (see Fig. 2 (a)). Films grown at 160ºC often showed a mixed growth mode, with columns and micro-crystallites either perpendicular or parallel to a surface.

Crystallinity of ZnCoO films was investigated with X-ray Diffraction (XRD). The XRD investigations showed that ZnCoO films are polycrystalline with wurtzite structure. In Fig. 3 we compare the XRD spectra of as-grown sample (grown at 160ºC) and the same sample after annealing at 800°C. No significant changes were observed after annealing at lower temperatures, but after annealing at 800°C the XRD peaks shift towards larger angles indicating a decrease of lattice constants. Origin of this decrease will be discussed in the next part of this paper. We observed also improvement of sample crystallinity after annealing. The relevant XRD peaks are sharper and more intense, as shown in Fig. 3.

The crucial question is whether Co ions substitute for $Zn^{2+}$ in the ZnO lattice in films grown at so low temperatures We looked for intra-shell transitions of Co ions in 2+ charge state. Characteristic features of d-d intra-shell transitions of $Co^{2+}$ ions from the $^4A_2$ ground state to the excited $^2E$ $(G)$ and $^4T_1$ $(P)$ states, (see reference [13] for details) were observed by us in optical transmission spectra at about 1.8 – 1.9 eV, indicated with arrows in Fig. 4. These investigations were performed for the 2850 nm thick as-grown $Zn_{0,99}Co_{0,01}O$ film. We thus claim that Co ions substitute $Zn^{2+}$ cations, since in this case Co will be in 2+ charge state.

Co introduction to the films was also confirmed by magnetic investigations with Superconducting Quantum Interference Device (SQIUD). In accordance with the theoretical calculations of Dietl et al. [6] to achieve ferromagnetic phase at RT Mn in ZnO should remain in 2+ charge state in p-type samples [6]. We thus checked how it is in the case of Co ions. An additional absorption band is observed in Co doped ZnO films, starting at about 2.5 eV. This band, which is Co related as al already reported in the reference [13], we attribute to a charge-transfer transition of Co ions. If this attribution is correct Co may change its charge state in the ZnO lattice.

Here we assume 2+ to 3+ ionization, since Co remained in the 2+ state in strongly n-type samples, but formation of so-called charge transfer state is also possible. Consequences of Co (and also Mn [14]) recharging in ZnO for magnetic ordering in ZnTMO are still not clear. It may depend on origin of recharging process, as discussed recently by us for ZnMnO [15] and also discussed in new theoretical models presented in the references [16,17]. We underline that recharging does not mean that strong magnetic effects will not be present. For example, giant magneto-optical effects were reported for p-type GaMnN films [18], in which Mn was in 3+ charge state and not 2+, as in the original theoretical model.

Introduction of Co to ZnO quenches visible PL/CL. This effect is clearly Co related, as we concluded comparing emission intensity in samples with different Co fractions. No visible PL or CL (see Fig. 5) was observed in samples showing uniform Co distribution and Co fraction above 1 atomic %.

The post-growth annealing at 800°C helps to recover visible emission. This is observed together with the appearance of strong green emission and a decrease of the lattice constant (observed in the XRD). Likely, Co out-diffuses from lattice sites, which results in some Co-deficient regions in the sample, from which a visible PL/CL emission is observed.

Following ref. 12 two conditions turned out to be important to obtain uniform ZnCoO films. First, the films should be grown at low temperature (about 160°C). Second, pulses with Co precursors should be separated by several pulses with zinc precursor.

Magnetic properties of ZnCoO films, as assessed by the SQUID magnetometry, confirm the above finding. Films with uniform Co distribution and with Co fraction below 5% show paramagnetic properties. Less uniform samples grown at the temperatures above 250ºC show a weak ferromagnetic response at low temperatures, as exemplified in Fig. 6. Importantly, the post-growth annealing, used by us for improvement of films crystallinity, does not affect magnetic properties of the films grown at LT. We observed only some Co out-diffusion in the case of the highest annealing temperature.

The electrical measurements were conducted at room temperature on the ZnCoO samples grown on a glass substrate. These measurements show that ZnCoO films are of n-type, with free carrier concentration varied between $10^{17}$ cm$^{-3}$ and $10^{19}$ cm$^{-3}$, depending on growth conditions. Free electron concentration depended mostly on the growth temperature, Zn-to-Co ratio and length of precursors pulses used in the ALD process, and not Co fraction.

## 4. Conclusions

ZnCoO films were grown at low temperature by the Atomic Layer Deposition. These films show very uniform Co distribution. In this case only a paramagnetic response is detected. Weak ferromagnetic phase was found in nonuniform samples, the one grown at higher temperature. This allows us to claim that paramagnetic response is due to high uniformity of Co distribution, and that ferromagnetic response is related to formation of some foreign chemical phases or due to Co accumulation. Post-growth annealing at high temperatures results in a better crystallinity of the films, but affects Co concentration in the films. The present results suggest Co out-diffusion from the annealed samples. Importantly, such annealing (improving sample crystallinity) is not affecting magnetic properties of the investigated films.

**Acknowledgements**

This work was supported by FunDMS ERC Advanced Grant Reserch.

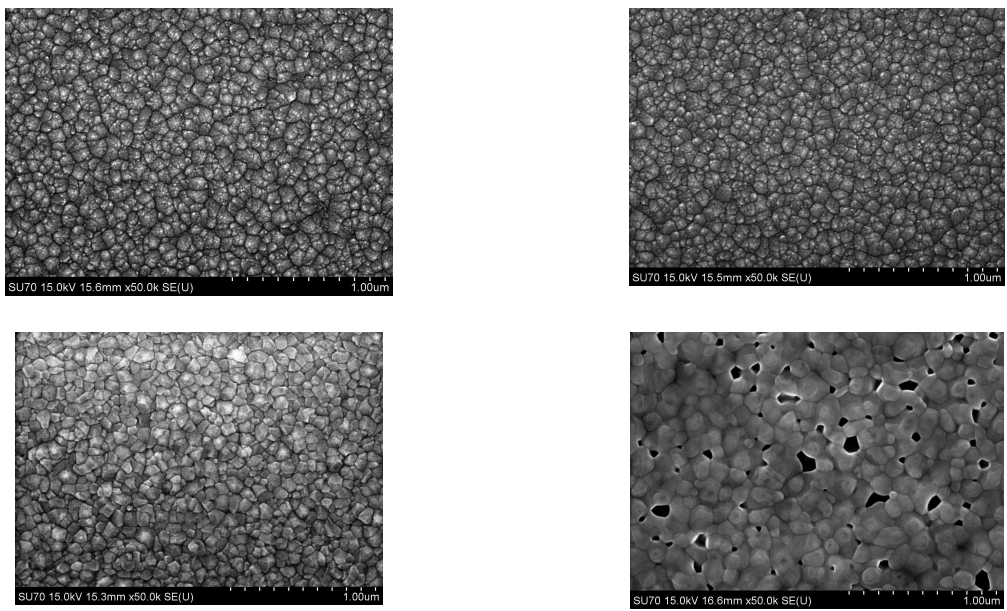

**Figure 1.** SEM images of the ZnCoO (2% of Co) film grown by the ALD on silicon substrate for: (a) as-grown sample (grown at 160°C) and after annealing of this sample at: (b) at 400°C, (c) at 600°C, and (d) at 800°C.

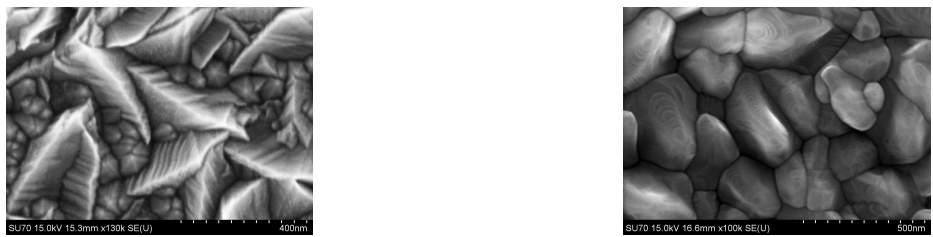

**Figure 2.** SEM images of ZnCoO film for: (a) as-grown sample (grown at 160ºC), and (b) the same sample annealed at 800°C.

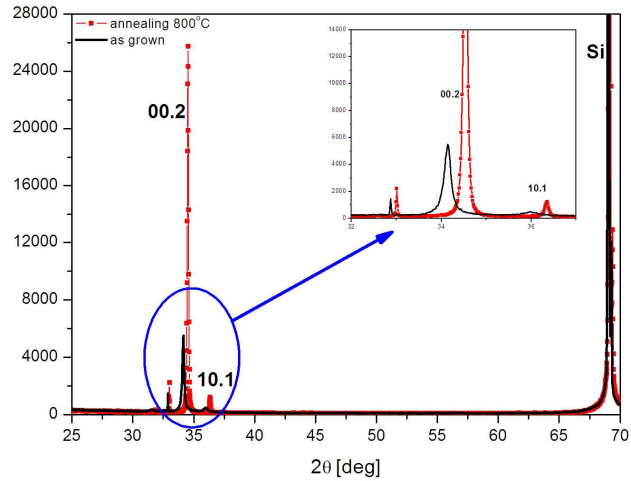

**Figure 3.** X-ray diffraction (XRD) spectra of as-grown ZnCoO film (grown by the ALD at 160°C) and after annealing at 800°C (red color online). 00.2 and 10.1 XRD reflections are shown.

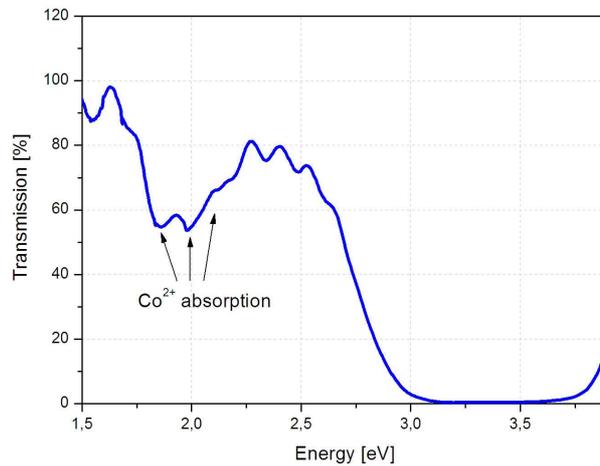

**Figure 4.** Transmission spectrum of 2850 nm thick $Zn_{0.99}Co_{0.01}O$ film grown by the ALD.

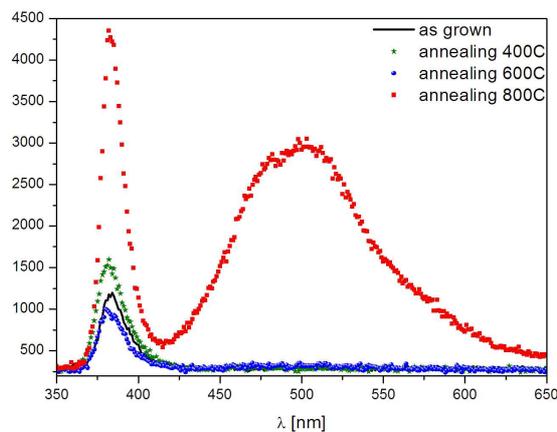

**Figure 5.** Room temperature CL spectra for ZnCoO thin film grown at 160°C on silicon substrate (color online).

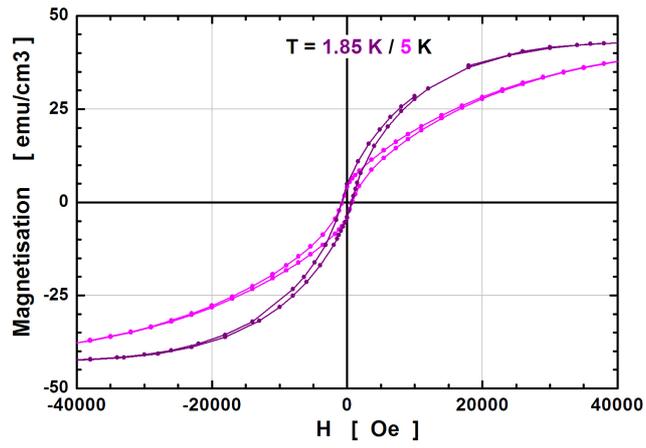

**Figure 6.** The M-H curves of ZnCoO film grown by the ALD at 300ºC with a Co concentration of 5% (color online).